\documentclass[aps,prd,preprint,superscriptaddress,tightenlines,nofootinbib]{revtex4}

\usepackage{amsfonts,amsmath,amsthm,amssymb}
\usepackage{mathrsfs}
\usepackage{bm}
\usepackage{color}
\usepackage{epsfig}

\newcommand{\PRE}[1]{{#1}}   

\newcommand{\postscript}[2]{\setlength{\epsfxsize}{#2\hsize}
   \centerline{\epsfbox{#1}}}
\newcommand{\comment}[1]{}

\begin{document}

\newcommand{\hhat}[1]{\hat {\hat{#1}}}
\newcommand{\pslash}[1]{#1\llap{\sl/}}
\newcommand{\kslash}[1]{\rlap{\sl/}#1}
\newcommand{\qq}{$^\sharp$ }
\newcommand{\rmb}{{\color{red}$^\bigstar$} }
\newcommand{\fp}{{\color{blue}$^\heartsuit$} }
\newcommand{\qa}{$^\mho$ }
\newcommand{\rp}{{\color{magenta}$^\Re$} }
\newcommand{\lab}[1]{\hypertarget{lb:#1}}
\newcommand{\iref}[2]{\footnote{\hyperlink{lb:#1}{\textit{$^\spadesuit$#2}}}}
\newcommand{\emp}[1]{{\bf\color{red} #1}}
\newcommand{\eml}[1]{{\color{blue} #1}}
\newcommand{\kw}[1]{\emph{#1}}
\newcommand{\sos}[1]{{\large \textbf{#1}}}
\newcommand{\soso}[1]{\chapter{#1}}
\newcommand{\sossub}[1]{\section{#1}}
\newcommand{\et}[1]{{\large\begin{flushleft} \color{blue}\textbf{#1} \end{flushleft}}}
\newcommand{\sto}[1]{\begin{center} \textit{#1} \end{center}}
\newcommand{\rf}[1]{{\color{blue}[\textit{#1}]}}

\newcommand{\el}[1]{\label{#1}}
\newcommand{\er}[1]{\eqref{#1}}
\newcommand{\df}[1]{\textbf{#1}}
\newcommand{\mdf}[1]{\pmb{#1}}
\newcommand{\ft}[1]{\footnote{#1}}
\newcommand{\n}[1]{$#1$}
\newcommand{\fals}[1]{$^\times$ #1}
\newcommand{\new}{{\color{red}$^{NEW}$ }}
\newcommand{\ci}[1]{}

\newcommand{\de}[1]{{\color{green}\underline{#1}}}
\newcommand{\ke}{\rangle}
\newcommand{\br}{\langle}
\newcommand{\lb}{\left(}
\newcommand{\rb}{\right)}
\newcommand{\rc}{\right.}
\newcommand{\lsb}{\left[}
\newcommand{\rsb}{\right]}
\newcommand{\blb}{\Big(}
\newcommand{\brb}{\Big)}
\newcommand{\nn}{\nonumber \\}
\newcommand{\p}{\partial}
\newcommand{\pd}[1]{\frac {\partial} {\partial #1}}
\newcommand{\cd}{\nabla}
\newcommand{\cc}{$>$}
\newcommand{\ba}{\begin{eqnarray}}
\newcommand{\ea}{\end{eqnarray}}
\newcommand{\be}{\begin{equation}}
\newcommand{\ee}{\end{equation}}
\newcommand{\bay}[1]{\left(\begin{array}{#1}}
\newcommand{\eay}{\end{array}\right)}
\newcommand{\eg}{\textit{e.g.} }
\newcommand{\ie}{\textit{i.e.}, }
\newcommand{\iv}[1]{{#1}^{-1}}
\newcommand{\st}[1]{|#1\ke}
\newcommand{\at}[1]{{\Big|}_{#1}}
\newcommand{\zt}[1]{\rm{#1}}
\newcommand{\zi}[1]{\textit{#1}}
\def\xa{{\alpha}}
\def\xA{{\Alpha}}
\def\xb{{\beta}}
\def\xB{{\Beta}}
\def\xd{{\delta}}
\def\xD{{\Delta}}
\def\xe{{\epsilon}}
\def\xE{{\Epsilon}}
\def\xve{{\varepsilon}}
\def\xg{{\gamma}}
\def\xG{{\Gamma}}
\def\xk{{\kappa}}
\def\xK{{\Kappa}}
\def\xl{{\lambda}}
\def\xL{{\Lambda}}
\def\xo{{\omega}}
\def\xO{{\Omega}}
\def\xvp{{\varphi}}
\def\xs{{\sigma}}
\def\xS{{\Sigma}}
\def\xt{{\theta}}
\def\xT{{\Theta}}
\def \Tr {{\rm Tr}}
\def\CA{{\cal A}}
\def\CC{{\cal C}}
\def\CD{{\cal D}}
\def\CE{{\cal E}}
\def\CF{{\cal F}}
\def\CH{{\cal H}}
\def\CJ{{\cal J}}
\def\CK{{\cal K}}
\def\CL{{\cal L}}
\def\CM{{\cal M}}
\def\CN{{\cal N}}
\def\CO{{\cal O}}
\def\CP{{\cal P}}
\def\CQ{{\cal Q}}
\def\CR{{\cal R}}
\def\CS{{\cal S}}
\def\CT{{\cal T}}
\def\CV{{\cal V}}
\def\CW{{\cal W}}
\def\CY{{\cal Y}}
\def\CZ{{\cal Z}}
\def\BC{\mathbb{C}}
\def\BR{\mathbb{R}}
\def\BZ{\mathbb{Z}}
\def\sA{\mathscr{A}}
\def\sB{\mathscr{B}}
\def\sD{\mathscr{D}}
\def\sE{\mathscr{E}}
\def\sF{\mathscr{F}}
\def\sG{\mathscr{G}}
\def\sH{\mathscr{H}}
\def\sJ{\mathscr{J}}
\def\sL{\mathscr{L}}
\def\sM{\mathscr{M}}
\def\sN{\mathscr{N}}
\def\sO{\mathscr{O}}
\def\sP{\mathscr{P}}
\def\sR{\mathscr{R}}
\def\sQ{\mathscr{Q}}
\def\sX{\mathscr{X}}

\def\slz{SL(2,\BZ)}
\def\slr{$SL(2,R)\times SL(2,R)$ }
\def\ads{${AdS}_5\times {S}^5$ }
\def\adst{${AdS}_3$ }
\def\sun{SU(N)}
\def\ad#1#2{{\frac \delta {\delta\sigma^{#1}} (#2)}}
\def\bqf{\bar Q_{\bar f}}
\def\nf{N_f}
\def\sunf{SU(N_f)}
\def\dcirc{{^\circ_\circ}}
\def\btr{{B_\nu{}^\nu}}
\def\byy{{B_{yy}}}
\def\sy{{\zt{sgn}(y)}}

\preprint{
\hfil
\begin{minipage}[t]{3in}
\begin{flushright}
\vspace*{.4in}
\end{flushright}
\end{minipage}
}

\vspace{1cm}

\title{Searching for string resonances in $\bm{e^+e^-}$ and $\bm{\gamma\gamma}$ collisions
\PRE{\vspace*{0.3in}} }

\author{Luis A. Anchordoqui}
\affiliation{Department of Physics,\\
University of Wisconsin-Milwaukee,
 Milwaukee, WI 53201, USA
\PRE{\vspace*{.1in}}
}

\author{Wan-Zhe Feng}
\affiliation{Department of Physics,\\
Northeastern University, Boston, MA 02115, USA
\PRE{\vspace*{.1in}}
}

\author{Haim \nolinebreak Goldberg}
\affiliation{Department of Physics,\\
Northeastern University, Boston, MA 02115, USA
\PRE{\vspace*{.1in}}
}

\author{Xing Huang}
\affiliation{Department of Physics,\\
University of Wisconsin-Milwaukee,
 Milwaukee, WI 53201, USA
\PRE{\vspace*{.1in}}
}

\author{Tomasz R. Taylor}
\affiliation{Department of Physics,\\
  Northeastern University, Boston, MA 02115, USA \PRE{\vspace*{.1in}}
}

\date{December 2010}

\PRE{\vspace*{.5in}}

\begin{abstract}\vskip 3mm
  \noindent If the fundamental mass scale of superstring theory is as
  low as few TeVs, the massive modes of vibrating strings, Regge
  excitations, will be copiously produced at the Large Hadron Collider
  (LHC).  We discuss the complementary signals of low mass
  superstrings at the proposed electron-positron facility (CLIC), in
  $e^+e^-$ and $\gamma\gamma$ collisions.  We examine all relevant
  four-particle amplitudes evaluated at the center of mass energies
  near the mass of lightest Regge excitations and extract the
  corresponding pole terms.  The Regge poles of {\em all\/} four-point
  amplitudes, in particular the spin content of the resonances, are
  completely model independent, universal properties of the entire
  landscape of string compactifications. We show that $\gamma \gamma
  \to e^+ e^-$ scattering proceeds only through a spin-2 Regge
  state. We estimate that for this particular channel, string scales
  as high as 4~TeV can be discovered at the 11$\sigma$ level with the
  first fb$^{-1}$ of data collected at a center-of-mass energy
  $\approx 5$~TeV. We also show that for $e^+e^-$ annihilation into
  fermion-antifermion pairs, string theory predicts the {\em precise}
  value, equal 1/3, of the relative weight of spin 2 and spin 1
  contributions. This yields a dimuon angular distribution with a
  pronounced forward-backward asymmetry, which will help
  distinguishing between low mass strings and other beyond the
  standard model scenarios.
\end{abstract}

\maketitle

\section{Introduction}

$e^+ e^-$ linear colliders are considered as the most desirable
facility to complement measurements at the Large Hadron Collider
(LHC). Two alternative linear projects are presently under
consideration: the International Linear Collider (ILC) and the Compact
LInear Collider (CLIC). The first one is based on superconducting
technology in the TeV range, whereas the second one is based on the
novel approach of two beam acceleration to extend linear colliders
into the multi-TeV range. The choice will be based on the respective
maturity of each technology and on the physics requests derived from
the LHC physics results when available.

CLIC aims at multi-TeV collision energy with high-luminosity, ${\cal
  L}_{e^+e^-} \sim 8 \times 10^{34}~{\rm cm}^{-2} \, {\rm
  s}^{-1}$~\cite{Ellis:2000iw}. The facility would be built in
phases. The initial center-of-mass energy has been arbitrarily chosen
to be $\sqrt{s} = 500$~GeV to allow direct comparison with ILC. The
collider design has been optimized for $\sqrt{s} = 3$~TeV, with a
possible upgrade path to $\sqrt{s} = 5$~TeV at constant
luminosity~\cite{Accomando:2004sz}. To keep the length (and thereby
the cost) of the machine at a reasonable level, the CLIC study
foresees a two beam accelerating scheme featuring an accelerating
gradient in the presence of a beam (loaded) in the order of 80~MV/m
and 100~MV/m, for the 500~GeV and 3~TeV options; the projected total
site lengths are 13.0~km and 48.3~km,
respectively~\cite{Ellis:2008gj}. The CLIC technology is less mature
than that of the ILC. In particular, the target accelerating gradient
is considerable higher than the ILC and requires very aggressive
performance from accelerating structures.

In addition, photon collisions that will considerably enrich the CLIC
physics program can be obtained for a relatively small incremental
cost. Recently, an exploratory study has been carried out to determine
how this facility could be turned into a collider with a high
geometric luminosity, which could be used as the basis for a $\gamma
\gamma$ collider~\cite{Telnov:2009vq}. The hard photon beam of the $\gamma\gamma$ collider
can be obtained by using the laser back-scattering technique, i.e.,
the Compton scattering of laser light on the high energy
electrons~\cite{Ginzburg:1981ik}. The scattered photons have energies
close to the energy of the initial electron beams, and the expected
$\gamma \gamma$ and $\gamma e$ luminosities can be comparable to that
in $e^+e^-$ collisions, {\em e.g.,} ${\cal L}_{\gamma \gamma} \sim 2
\times 10^{34}~{\rm cm}^{-2} \, {\rm s}^{-1}$.

If either supersymmetry (SUSY) or extra dimensions exist at the TeV
scale, signals of new physics should be found at the LHC. However, the
proper interpretation of such discoveries, namely the correct
identification and the nature of the new physics signals, may not be
straightforward at the LHC and may require complementary data from
CLIC. In particular, a multi-TeV collider would ensure a sensitivity
over a broad mass range allowing a complete investigation of the SUSY
particle spectrum~\cite{Battaglia:2002eg}.  Alternatively, distinct
signals of new vector resonances and quantum black holes could also be
at reach~\cite{Battaglia:2003cn}. Along the lines, in this work we
discuss direct searches of string physics at CLIC drawing upon LHC
techniques developed
elsewhere~\cite{Anchordoqui:2007da,Anchordoqui:2008ac,Anchordoqui:2008hi,Lust:2008qc,Anchordoqui:2008di,Lust:2009pz,Anchordoqui:2009mm,Anchordoqui:2009ja,Feng:2010yx}.

In string theory, elementary particles are quantized vibrations of
fundamental strings. The zero modes are massless, the first harmonics
have masses equal to the fundamental mass $M$, the second $\sqrt{2}M$
and, in general
\begin{equation} M_n=\sqrt{n}M\ .\end{equation} These massive {\em
  Regge\/} particles have higher spins, ranging from 0 to $n{+}1$ and
come in $SU(3)\times SU(1)\times U(1)_Y$ representations copied from
gauge bosons, quarks and leptons. For example, gluon's lowest Regge
excitations are spin 0, 1 and 2 color octets. The Standard Model (SM)
spectrum is replicated at mass $M$ and then at each $\sqrt{n}M$
level. It is possible that loop corrections can split some levels,
however this infinite replication is the most fundamental property of
string theory.

If, as commonly believed, $M$ is in the Planckian regime, then the
landscape problem makes it very difficult to connect string theory to
experimental data.  However theoretically, $M$ can be as low as few
TeVs, provided that Nature endowed us with some large extra
dimensions, with typical length scale of order
0.1~mm~\cite{Antoniadis:1998ig}.  Such a ``low string mass'' scenario
leads to some spectacular experimental consequences, universal to all
compactifications thus insensitive to the landscape problem. After
operating for only few months, with merely 2.9 inverse picobarns of
integrated luminosity, the LHC CMS experiment has recently ruled out
$M <2.5$~TeV by searching for narrow resonances in the dijet mass
spectrum~\cite{Khachatryan:2010jd}. In fact, LHC has the capacity of
discovering strongly interacting resonances in practically all range
up to $\sqrt{s}_{\rm LHC}$. The present study is based on the
optimistic assumption that by the time the ILC/CLIC start operating,
there will be at least some indications for the existence of Regge
resonances. We will argue that the proposed $e^+e^-$ and
$\gamma\gamma$ colliders offer an excellent opportunity for probing
string physics.

The layout of the paper is as follows. In Sec.~\ref{II} we outline the basic setting of TeV-scale  string compactifications and discuss general aspects of  intersecting D-brane configurations that realize the SM  by open strings. In Sec.~\ref{III}  we present a complete calculation of all relevant four-point string scattering amplitudes. The computation  is performed in a model independent and universal way, and so our results hold for all compactifications. In Sec.~\ref{IV}  we discuss the associated phenomenological aspects of  Regge recurrences of open strings related to experimental searches for new physics at  CLIC. Our conclusions are collected in Sec.~\ref{V}.

\section{Photon in the intersecting brane SM constructions}
\label{II}

TeV-scale superstring theory provides a brane-world description of the  SM, which is localized on D-branes extending in $p+3$ spatial dimensions. Gauge interactions emerge as excitations of open strings with endpoints attached on the D-branes, whereas gravitational interactions are described by closed strings that can propagate in all nine spatial dimensions of string theory (these comprise parallel dimensions extended along the $(p+3)$-branes and transverse dimensions).

The basic unit of gauge invariance for D-brane constructions is a $U(1)$ field, and so one can stack up $N$ identical D-branes to generate a $U(N)$ theory with the associated $U(N)$ gauge group.  Gauge bosons are due to strings attached to stacks of D-branes and chiral matter due to strings stretching between intersecting D-branes~\cite{Blumenhagen:2006ci}.  Each of the two strings endpoints carries a fundamental charge with respect to the stack of branes on which it terminates.  Mater fields carry quantum numbers associated with bifundamental representations.

While the existence of Regge excitations is a completely universal
feature of string theory, there are many ways of realizing SM in such
a framework.  Individual models utilize various D-brane configurations
and compactification spaces. They may lead to very different SM
extensions, but as far as the collider signatures of Regge excitations
are concerned, their differences boil down to a few parameters. The
most relevant characteristics is how the $U(1)_Y$ hypercharge is
embedded in the $U(1)$s associated to $D$-branes. One $U(1)$ (baryon
number) comes from the ``QCD'' stack of three branes, as a subgroup of
the $U(3)$ group that contains $SU(3)$ color but obviously, one needs
at least one extra $U(1)$. In D-brane compactifications, hypercharge
always appears as a linear, non-anomalous combination of the baryon
number with one, two or more $U(1)$s. The precise form of this
combination bears down on the photon couplings, however the
differences between individual models amount to numerical values of a
few parameters.  In order to develop our program in the simplest way,
we work within the construct of a minimal model in which the color
stack $a$ of three D-branes are intersected by the (weak doublet)
stack $b$ and by one (weak singlet) D-brane
$c$~\cite{Antoniadis:2000ena}. {}For the two-brane stack $b$, there is
a freedom of choosing physical state projections leading either to
$U(2)_b$ or to the symplectic $Sp(1)$ representation of Weinberg-Salam
$SU(2)_L$~\cite{Berenstein:2006pk}.

In the bosonic sector, the open strings terminating on QCD stack $a$ contain the standard $SU(3)$ octet of gluons $g_\mu^a$ and an additional $U(1)_a$ gauge boson $C_\mu$, most simply the manifestation of a gauged baryon number symmetry: $U(3)_a\sim SU(3)\times U(1)_a$. On the $U(2)_b$ stack the open strings correspond to the electroweak gauge bosons $A_\mu^a$, and again an additional $U(1)_b$ gauge field $X_\mu$.  So the associated gauge groups for these stacks are $SU(3) \times U(1)_a,$ $SU(2)_L \times U(1)_b$, and $U(1)_c$, respectively.  We can further simplify the model by eliminating $X_\mu$; to this end instead we can choose the projections leading to $Sp(1)$ instead of $U(2)_b$ \cite{Berenstein:2006pk}. The $U(1)_Y$ boson $Y_\mu$, which gauges the usual electroweak hypercharge symmetry, is a linear combination of $C_\mu$, the $U(1)_c$ boson $B_\mu$, and perhaps a third additional $U(1)$ gauge field, $X_\mu$.  The fermionic matter consists of open strings located at the intersection points of the three stacks.  Concretely, the left-handed quarks are sitting at the intersection of the $a$ and the $b$ stacks, whereas the right-handed $u$ quarks comes from the intersection of the $a$ and $c$ stacks and the right-handed $d$ quarks are situated at the intersection of the $a$ stack with the $c'$ (orientifold mirror) stack. All the scattering amplitudes between these SM particles, which we will need in the following, essentially only depend on the local intersection properties of these D-brane stacks~\cite{MarchesanoBuznego:2003hp}.

\begin{table}
\caption{Chiral fermion spectrum of the $U(3)_a \times Sp(1)_L \times U(1)_c$ D-brane model.}
\begin{tabular}{c|ccccc}
\hline
\hline
 Name &~~Representation~~& ~$Q_{U(3)}$~& ~$Q_{U(1)}$~ & ~$Q_Y$~ \\
\hline
~~$U_i$~~ & $({\bar 3},1)$ &    $-1$ & $\phantom{-}1$ & $-\frac{2}{3}$ \\[1mm]
~~$D_i$~~ &  $({\bar 3},1)$ &    $-1$ & $-1$ & $\phantom{-}\frac{1}{3}$  \\[1mm]
~~$L_i$~~ & $(1,2)$&    $\phantom{-}0$ & $\phantom{-}1$ & $-\frac{1}{2}$  \\[1mm]
~~$E_i$~~ &  $(1,1)$&  $\phantom{-}0$ & $-2$ &  $\phantom{-}1$  \\[1mm]
~~$Q_i$~~ & $(3,2)$& $\phantom{-}1$ & $\phantom{-}0$ & $\phantom{-}\frac{1}{6}$ \\[1mm]
\hline
\hline
\end{tabular}
\label{t1}
\end{table}

The chiral fermion spectrum of the $U(3)_a \times Sp(1) \times U(1)_c$ D-brane model is given in Table~\ref{t1}.  In such a minimal D-brane construction, if the coupling strength of $C_\mu$ is down by root six when compared to the $SU(3)_C$ coupling $g_a$, the hypercharge $Q_Y\equiv \frac{1}{6} Q_{U(3)}-\frac{1}{2}Q_{U(1)}$ is free of anomalies. However, the $Q_{U(3)}$ (gauged baryon number) is anomalous.  This anomaly is canceled by the f-D version~\cite{Witten:1984dg} of the Green-Schwarz mechanism~\cite{Green:1984sg}.  The vector boson $Y'_\mu$, orthogonal to the hypercharge, must grow a mass in order to avoid long range forces between baryons other than gravity and Coulomb forces. The anomalous mass growth allows the survival of global baryon number conservation, preventing fast proton decay~\cite{Ghilencea:2002da}.

In the $U(3)_a \times Sp(1)_L \times U(1)_c$ D-brane model, the $U(1)_a$ assignments are fixed (they
give the baryon number) and the hypercharge assignments are fixed by SM. Therefore, the mixing angle $\theta_P$ between the hypercharge and the $U(1)_a$
is obtained in a similar manner to the way the Weinberg angle is fixed
by the $SU(2)_L$ and the $U(1)_Y$ couplings ($g_b$ and $g_Y$, respectively) in the SM. The Lagrangian containing the $U(1)_a$ and $U(1)_c$ gauge fields is given by
\begin{equation}
{\cal L} = g_c \, \hat B_\mu \, J_B^\mu + \frac{g_a}{\sqrt{6}} \, \hat  C_\mu  \,  J_C^\mu
\label{apache}
\end{equation}
where $\hat B_\mu = \cos \theta_P \, Y_\mu + \sin \theta_P \, Y'_\mu$
and
$\hat C_\mu = -\sin \theta_P \, Y_\mu + \cos \theta_P\, Y'_\mu$
are canonically normalized, and $g_c$ is the coupling strength of the $U(1)_c$ gauge field.  Substitution of these expressions into (\ref{apache}) leads to
\begin{eqnarray}
{\cal L}  =  Y_\mu \left(g_c \cos \theta_P J_B^\mu - \frac{g_a}{\sqrt{6}} \sin \theta_P J_C^\mu \right)
+  Y'_\mu \left(g_c \sin \theta_P J_B^\mu + \frac{g_a}{\sqrt{6}} \cos \theta_P J_C^\mu \right),
\end{eqnarray}
with $g_c \, \cos \theta_P\,  J_B^\mu - \frac{1}{\sqrt{6}}\,  g_a \, \sin \theta_P \,  J_C^\mu = g_Y \, J_Y^\mu$. We have seen
that the hypercharge is anomaly free if
$J_Y =  \frac{1}{6} \,  J_C^\mu - \frac{1}{2} \,  J_B^\mu$, yielding
\begin{equation}
g_c \cos \theta_P = \frac{1}{2} g_Y \quad {\rm and}  \quad\frac{g_a}{\sqrt{6}} \sin \theta_P = \frac{1}{6} g_Y \, .
\label{pipita}
\end{equation}
{}From (\ref{pipita}) we obtain the following relations
\begin{equation}
\tan \theta_P = \sqrt{\frac{2}{3}} \, \frac{g_c}{g_a}, \quad \quad
\left(\frac{g_Y}{2g_c}\right)^2 + \left(\frac{1}{\sqrt{6}} \frac{g_Y}{g_a}\right)^2 = 1, \quad {\rm and}
\quad \frac{1}{4 g_c^2} + \frac{1}{6 g_a^2} = \frac{1}{g_Y^2} \, .
\label{pellerano}
\end{equation}
We use the evolution of gauge couplings from the weak scale $M_Z$ as determined by the one-loop beta-functions of the SM with three families of quarks and leptons and one Higgs doublet,
\begin{equation}
{1\over \alpha_i(M)}={1\over \alpha_i(M_Z)}-
{b_i\over 2\pi}\ln{M \over M_Z}\ ; \quad i=a,b,Y,
\end{equation}
where $\alpha_i=g_i^2/4\pi$ and $b_a=-7$, $b_b=-19/6$, $b_Y=41/6$. We also use the measured values of the couplings at the $Z$ pole $\alpha_a(M_Z)=0.118\pm 0.003$, $\alpha_b(M_Z)=0.0338$, $\alpha_Y(M_Z)=0.01014$ (with the errors in $\alpha_{b,Y}$ less than 1\%)~\cite{Amsler:2008zzb}.  Running couplings up to 3~TeV, which is where the phenomenology will be, we get $\kappa \equiv \sin \theta_P \sim 0.14$. When the theory undergoes electroweak symmetry breaking, because $Y'$ couples to the Higgs, one gets additional mixing. Hence $Y'$ is not exactly a mass eigenstate. The explicit form of the low energy eigenstates $A_\mu$, $Z_\mu,$ and $Z'_\mu$ is given in~\cite{Berenstein:2008xg}.

In the $U(3)_a \times U(2)_b \times U(1)_c$ D-brane model, the hypercharge is given by
\be\el{hyperchargeY} Q_Y = c_a Q_{U(3)} + c_b Q_{U(2)} + c_c Q_{U(1)}.\ee
Note that we have, in the covariant derivative $\CD_\mu$,
\be\el{covderi} \CD_\mu = \p_\mu -i g_c \, B_\mu \, Q_{U(1)}  -i \frac{g_b}{2} \, X_\mu \, Q_{U(2)}  - i \frac{g_a}{\sqrt{6}} \, C_\mu  \,  Q_{U(3)}.\ee
We can define $Y_\mu$ and two other fields $Y'{}_\mu, Y''{}_\mu$ that are related to $C_\mu, X_\mu, B_\mu$ by a orthogonal transformation $O$ defined as
\[\bay{c} Y \\ Y' \\ Y''\eay = O \bay{c} C \\ X \\ B\eay.\]
In order for $Y_\mu$ to have the hypercharge $Q_Y$ as in
Eq.~\er{hyperchargeY}, we need, \be\el{othogaugefield} C_\mu = \frac
{\sqrt{6}c_a g_Y} {g_a} Y_\mu + \dots,\quad X_\mu = \frac {2 c_b g_Y}
{g_b} Y_\mu + \dots,\quad B_\mu =\frac {c_c g_Y} {g_c} Y_\mu +
\dots.\ee where $g_Y$ is given by \be \frac{1}{g_Y^2} = \frac{6
  c_a^2}{ g_a^2} + \frac{4 c_b^2}{g_b^2} + \frac{c_c^2}{g_c^2}.\ee The
field $Y_\mu$ then appears in the covariant derivative with the
desired $Q_Y$, \be\el{covderiY}\CD_\mu = \p_\mu -i g_Y Y_\mu Q_Y +
\dots.\ee The ratio of the coefficients in Eq.~\er{othogaugefield} is
determined by the form of Eq.~\er{hyperchargeY} and
Eq.~\er{covderi}. More explicitly, only with such ratio, we can have
$Q_Y$ in Eq.~\er{covderiY}. The value of $g_Y$ is determined so that
the coefficients in Eq.~\er{othogaugefield} are components of a
normalized vector so that they can be a row vector of $O$. The rest of
the transformation (the ellipsis part) involving $Y',Y''$ is not
necessary for our calculation. The point is that we now know the first
row of the matrix $O$ and hence we can get the first column of $O^T$,
which gives the expression of $Y_\mu$ in terms of $C_\mu, X_\mu,
B_\mu$, \be Y_\mu = \frac {\sqrt{6}c_a g_Y} {g_a} C_\mu + \frac {2 c_b
  g_Y} {g_b} X_\mu +\frac {c_c g_Y} {g_c} B_\mu.\ee This is all we
need when we calculate the interaction involving $Y_\mu$; the rest of
$O$, which tells us the expression of $Y', Y''$ in terms of $C,X,B$ is
not necessary.  For later convenience, we define $\xk, \eta, \xi$ as
\be Y_\mu = \xk C_\mu + \eta X_\mu +\xi B_\mu\,;\ee therefore \be \xk
= \frac {\sqrt{6}c_a g_Y} {g_a},\quad \eta = \frac {2 c_b g_Y}
{g_b},\quad \xi = \frac {c_c g_Y} {g_c}.\label{consts}\ee

We pause to summarize the degree of model dependency stemming from the multiple $U(1)$ content of the minimal model containing 3 stacks of D-branes. First, there is an initial choice to be made for the gauge group living on the $b$ stack. This can be either $Sp(1)$ or $U(2)$. In the case of $Sp(1)$, the requirement that the hypercharge remain anomaly-free was sufficient to fix its $U(1)_a$ and $U(1)_c$ content, as explicitly presented in Eqs.~(\ref{pipita}) and (\ref{pellerano}). Consequently, the fermion couplings, as well as the mixing angle $\theta_P$ between hypercharge and the baryon number gauge field are wholly determined by the usual SM couplings.  The alternative selection -- that of $U(2)$ as the gauge group tied to the $b$ stack -- branches into some further choices. This is because the $Q_a,\ Q_b,\ Q_c$ content of the hypercharge operator is not uniquely determined by the anomaly cancelation requirement.  In fact, as seen in~\cite{Antoniadis:2000ena}, there are 5  possibilities. This final choice does not depend on further symmetry considerations; in Ref.~\cite{Antoniadis:2000ena} it was fixed ($c_a =2/3 ,\ c_b =1/2 ,\ c_c=1$) by requiring partial unification ($g_a = g_b$) and acceptable value of $\sin^2 \theta_W$ at string scales of 6 to 8 TeV. In Ref.~\cite{Anchordoqui:2011ag}, a different choice is made ($c_a = -2/3 , c_b = 1 , c_c = 0$ ) to  explain the CDF anomaly~\cite{Aaltonen:2011mk}. Clearly the mixing possibilities within the $U(1)_a\times U(1)_b\times U(1)_c$ serve to introduce a discrete number of phenomenological ambiguities. This contrasts strongly with the case where all the scattering evolves on one brane ({\em e.g.,} the $a$ stack on the color brane, which serves as the locale for stringy dijet processes at LHC.~\cite{Anchordoqui:2008di}).

In principle, in addition to the orthogonal field mixing induced by identifying anomalous and non-anomalous $U(1)$ sectors, there may be kinetic mixing between these sectors.  In our case, however, since there is only one $U(1)$ per stack of D-branes, the relevant
  kinetic mixing is between  $U(1)$'s  on different stacks, and hence involves
  loops with fermions at brane intersection.  Such loop  terms are typically
  down by $g_i^2/16 \pi^2 \sim 0.01$~\cite{Dienes:1996zr}. Generally, the major effect of the kinetic mixing is in communicating SUSY breaking from a hidden $U(1)$ sector to the visible sector, generally in modification of soft scalar masses.  Stability of the weak scale in various models of SUSY breaking requires the mixing to be orders of magnitude below these values~\cite{Dienes:1996zr}. For a comprehensive review of experimental limits on the mixing, see~\cite{Abel:2008ai}.  Moreover, the model discussed in the present work does not have a hidden sector-- all our $U(1)$'s (including the anomalous ones) couple to the visible sector.\footnote{We also work in the weak coupling regime. For an alternate  approach, see~\cite{Kitazawa:2009kr}.} In summary, kinetic mixing between the non-anomalous and the anomalous $U(1)$'s in our basic three stack model will be small because the fermions in the loop are all in the visible sector. In the absence of electroweak symmetry breaking, the mixing vanishes.

The scattering amplitudes involving four gauge bosons as well as those with two
gauge bosons plus two leptons do not depend on the compactification
details of the transverse space~\cite{Lust:2008qc}.\footnote{The only remnant
of the compactification is the relation between the Yang-Mills
coupling and the string coupling. We take this relation to reduce to
field theoretical results in the case where they exist, e.g., $gg \to
gg$. Then, because of the require correspondence with field theory,
the phenomenological results are independent of the compactification
of the transverse space. However, a different phenomenology would
result as a consequence of warping one or more parallel
dimensions~\cite{Hassanain:2009at}.} They will be particularly useful for testing low mass strings in $\gamma\gamma$ collisions. On the other hand,
the amplitudes involving four fermions, including $e^+e^-\to e^+e^-$, $e^+e^-\to \mu^+\mu^-$ and $e^+e^-\to q\bar q$ (in general, $e^+e^-\to F\bar F$, where $F\bar F$ is a fermion-antifermion pair), which are of particular interest for the
$e^+e^-$ collider, depend on the properties of extra dimensions and may include resonant contributions due to Kaluza-Klein excitations, string excitations of the Higgs scalar {\em etc}. However, it follows from Ref.\cite{Feng:2010yx} that the three-point couplings of Regge excitations to fermion-antifermion pairs are model-independent. Furthermore, the relative weights of resonances with different spins $J=0,1,2$ are unambigously predicted by the theory. Thus the resonant contributions to these amplitudes, with Regge excitations propagating in the $s$-channel,  are model-independent. $e^+e^-$ colliders can be used not only for discovering such resonances, but most importantly, for detailed studies of their spin content, therefore for distinguishing low mass string theory from other beyond the SM  extensions predicting the existence of similar particles.

\section{Regge resonances in $\bm{\gamma\gamma}$ and $\bm{e^+e^-}$ channels}
\label{III}
\subsection{Universal amplitudes for $\bm{\gamma\gamma}$ fusion}
\subsubsection{$\gamma\gamma\to \gamma\gamma$, $\gamma\gamma\to Z^0Z^0$, $\gamma\gamma\to W^+W^-$, $\gamma\gamma\to gg$}

As explained in the previous section, the electroweak hypercharge is a linear combination of charges associated  to different stacks of D-branes, therefore photons are linear combinations of three or more vector bosons. On the other hand, at the string disk level, non-vanishing amplitudes with no external particles other than gauge bosons always involve a single stack of D-branes at the disk boundary. Nevertheless, $\gamma\gamma$ fusion into gluon pairs {\em etc}.\ is possible already at this level because the two initial photons are superpositions of states associated to different stacks.
We will first study the resonant behavior of single-stack amplitudes and then compute the weights of the corresponding contributions to $\gamma\gamma$ processes under consideration.

All string disk amplitudes with four external gauge bosons $A$
can be obtained from the MHV amplitude \cite{Parke:1986gb}:\footnote{We use the standard notation of~\cite{Mangano:1990by}, although
the gauge group generators are normalized here in a different way,
according to ${\rm Tr}(T^{a}T^{b})=\frac{1}{2}\delta^{ab}$.}
 \begin{eqnarray}
{\cal M}(A_{1}^{-},A_{2}^{-},A_{3}^{+},A_{4}^{+}) & = & 4\, g^{2}\langle12\rangle^{4}\bigg[
\frac{V_{t}}{\langle12\rangle\langle23\rangle\langle34\rangle\langle41\rangle}
\makebox{Tr}(T^{a_{1}}T^{a_{2}}T^{a_{3}}T^{a_{4}}+T^{a_{2}}T^{a_{1}}T^{a_{4}}
T^{a_{3}})\nonumber \\
 & + & \frac{V_{u}}{\langle13\rangle\langle34\rangle\langle42\rangle\langle21\rangle}
 \makebox{Tr}(T^{a_{2}}T^{a_{1}}T^{a_{3}}T^{a_{4}}+T^{a_{1}}T^{a_{2}}T^{a_{4}}
 T^{a_{3}})\nonumber \\
 & + & \frac{V_{s}}{\langle14\rangle\langle42\rangle\langle23\rangle\langle31
 \rangle}\makebox{Tr}(T^{a_{1}}T^{a_{3}}T^{a_{2}}T^{a_{4}}+T^{a_{3}}T^{a_{1}}
 T^{a_{4}}T^{a_{2}})\bigg],\label{mhv}\end{eqnarray}
where the string {}``formfactor'' functions of the Mandelstam variables
$s,t,u~(s+t+u=0)$%
\footnote{Here, $s,t,u$ refer to parton subprocesses.} are defined as \begin{equation}
V_{t}=V(s,t,u)~,\qquad V_{u}=V(t,u,s)~,\qquad V_{s}=V(u,s,t)~,\end{equation}
with \begin{equation}
V(s,t,u)=\frac{s\, u}{tM^{2}}B(-s/M^{2},-u/M^{2})=\frac{\Gamma(1-s/M^{2})\ \Gamma(1-u/M^{2})}{\Gamma(1+t/M^{2})}.\label{formf}\end{equation}
The amplitudes have  $s$-channel poles at each $s=nM^2$, as seen from the expansion \cite{Veneziano:1968yb}:
\begin{equation}
B(-s/M,-u/M^{2})=-\sum_{n=0}^{\infty}\frac{M^{2-2n}}{n!}
\frac{1}{s-nM^{2}}\Bigg[\prod_{J=1}^{n}(u+M^{2}J)\Bigg],\label{bexp}
\end{equation}
 reflecting the propagation of resonances with spins up to $n+1$.

We first focus on the lowest, $n=1$ resonances. Near $s= M^{2}$,
$V_{s}$ is regular while \begin{equation}
V_{t}\to \frac{u}{s-M^{2}}~,\qquad V_{u}\to \frac{t}{s-M^{2}}~.\end{equation}
Thus the $s$-channel pole term of the amplitude (\ref{mhv}), relevant
to $(--)$ decays of intermediate states, is \begin{equation}
{\cal M}(A_{1}^{-},A_{2}^{-},A_{3}^{+},A_{4}^{+})\to2\, g^{2}\,{\cal C}^{1234}\frac{\langle12\rangle^{4}}{\langle12\rangle\langle23\rangle\langle34\rangle\langle41\rangle}\frac{u}{s-M^{2}}\ ,\label{mhvs}\end{equation}
where \begin{equation}
\CC^{1234}=2\Tr(\{T^{a_{1}},T^{a_{2}}\}\{T^{a_{3}},T^{a_{4}}\})=
16\sum_{a=0}^{N^{2}-1}d^{a_{1}a_{2}a}d^{a_{3}a_{4}a}\,.\label{groupfacs}\end{equation}
The amplitude with the $s$-channel pole relevant to $(+-)$ decays
is \begin{equation}
{\cal M}(A_{1}^{-},A_{2}^{+},A_{3}^{+},A_{4}^{-})\to2\, g^{2}\,{\cal C}^{1234}\frac{\langle14\rangle^{4}}{\langle12\rangle\langle23\rangle
\langle34\rangle\langle41\rangle}\frac{u}{s-M^{2}}\ .\label{mhvo}\end{equation}
\begin{table}
\caption{Group factors and couplings for the pole terms (\ref{mhvs}) and (\ref{mhvo}).}
\begin{tabular}{l|cc}
\hline
\hline
~~ Process~~& ~~Coupling~~ &~~~~$\CC^{1234}$~~~~ \\
\hline
$CC\to gg$ & $g_a^2$  & $\frac{2}{3}\delta_{a_{3}a_{4}}$ \\[1mm]
$CC\to CC$ &  $g_a^2$ & $\frac{2}{3}$ \\ \hline
$XX\to XX$ & $g_b^2 $& 1\\
$A^{3}A^{3}\to XX$ & $g_b^2$ & 1\\
$A^{3}A^{3}\to A^{3}A^{3}$ & $g_b^2$ & 1\\
$A^{3}X\to A^{3}X$ & $g_b^2$ & 1\\ \hline
$BB\to BB$& $2g_c^2$ & 2\\
\hline
\hline
\end{tabular}
\label{t2}
\end{table}
In Table~\ref{t2}, we list the group factors and couplings [replacing $g^2$ in Eqs.(\ref{mhvs}) and (\ref{mhvo})] for the single-stack processes contributing to $\gamma\gamma$ fusion into gauge bosons, evaluated according to Eq.(\ref{groupfacs}). \footnote{As can be seen in Eq.~(\ref{covderi}) the $X_\mu$ and $C_\mu$ normalization carries a factor $1/\sqrt{2N}$,  which is absent in the $B_\mu$ field. Hence, we should recover the $\sqrt{2N}$ factor (to be $B_\mu (\sqrt 2 g_c)/\sqrt 2 Q_{U(1)}$) and use $\sqrt 2 g_c$ in any calculation that follows from a general  $N$.}

We now proceed to higher level resonances, starting from
$n=2$. Three-particle amplitudes involving one level $n$ Regge
excitation (gauge index $a$) and two massless $U(N)$ gauge bosons
(gauge indices $a_1$ and $ a_2$) are even under the world-sheet parity
(reversing the order of Chan-Paton factors) for odd $n$, and odd for
even $n$ \cite{Feng:2010yx}. As a result, the respective group factors
are the symmetric traces $d^{a_1a_2a}$ for odd $n$ and non-abelian
structure constants $f^{a_1a_2a}$ for even $n$, respectively. For all
configuration of initial particles in the processes listed in
Table~\ref{t2}, $f^{a_1a_2a}=0$, therefore the corresponding
amplitudes have no $s$-channel poles associated to Regge resonances
with even $n$.\footnote{For $n=2$, this has already been checked by
  explicit computation in Ref.\cite{Dong:2010jt}.} {}For $USp(N)$
groups, the parity assignment is reversed, however the relevant
symmetric trace $d^{33a}=0$ for $Sp(1)$, therefore the same conclusion
holds for all SM embeddings under consideration. Thus in order to
observe higher level resonances, $\gamma\gamma$ collisions would have
to reach $\sqrt{s}>\sqrt{3}M$, which due to the recently established
$M>2.5$ TeV bound translates into $\sqrt{s}>4.3$ TeV. It is unlikely
that such high energies will be reached in the next generation of
$\gamma\gamma$ colliders, therefore from now on our discussion will be
limited to the lowest level resonances.

The $\gamma\gamma$ amplitudes are linear combinations of the amplitudes for processes listen in Table \ref{t2}, with the weights determined by the constants $\kappa$, $\eta$, $\xi$, {\em c.f}.\ Eq.(\ref{consts}), and the Weinberg angle $\theta_W$ with:
\be C_{W}=\cos\theta_{W} \quad,\quad S_{W}=\sin\theta_{W}.\ee
{}For the $U(3)_{a}\times U(2)_{b}\times U(1)_{c}$ minimal
model, they are given by:
\begin{eqnarray}
{\cal M} (\xg \xg \to gg)  & = & \kappa^2 C_W{}^2 \, {\cal M} (CC \to gg),  \\[1mm]
{\cal M} (\xg\xg \to \xg\xg) & = &  \xk^4 C_W{}^4 \, {\cal M} (C C \to CC) + 4 \eta^2 S_W{}^2 C_W{}^2 \, {\cal M} (X A^3 \to X A^3)  \nonumber \\
 & & +~\eta^4 C_W{}^4 \,  {\cal M} (X X \to XX)
 + S_W{}^4\,  {\cal M} (A^3 A^3 \to A^3 A^3)\nonumber  \\
& & +~\eta^2 S_W{}^2 C_W{}^2 \,  {\cal M} (A^3 A^3 \to XX) +~\eta^2 S_W{}^2 C_W{}^2 \,  {\cal M} (X X \to A^3 A^3)\nn
& & +~\xi^4 C_W{}^4 \,  {\cal M} (BB \to BB) \nonumber\\[1mm]
& = &  \xk^4 C_W{}^4 \, {\cal M} (C C \to CC) + 4 \eta^2 S_W{}^2 C_W{}^2 \, {\cal M} (X A^3 \to X A^3)  \nonumber \\
 & & +~(S_W{}^4+\eta^4 C_W{}^4 + 2\eta^2 S_W{}^2 C_W{}^2) \,  {\cal M} (X X \to XX)  \nonumber\\
& & +~\xi^4 C_W{}^4 \,  {\cal M} (BB \to BB) \,, \\
{\cal M} (\xg \xg \to Z^0Z^0)  & = &  \xk^4 C_W{}^2 S_W{}^2\, {\cal M} (C C \to CC) + 4 \eta^2 S_W{}^2 C_W{}^2 \, {\cal M} (X A^3 \to X A^3)  \nonumber \\
 & & +(S_W{}^2 C_W{}^2+\eta^4 C_W{}^2S_W{}^2 + \eta^2 S_W{}^4+\eta^2 C_W{}^4) \,  {\cal M} (X X \to XX) \nonumber \\
& & +~\xi^4 S_W{}^2 C_W{}^2  \,  {\cal M} (BB \to BB) \, ,\\
{\cal M} (\xg \xg \to W^+W^-)  & = & ~\eta^2 C_W{}^2 \,  {\cal M} (X X \to W^+W^-)
 + S_W{}^2\,  {\cal M} (A^3 A^3 \to W^+W^-)\nonumber  \nn
& = & ~(\eta^2 C_W{}^2+ S_W{}^2)\CM(XX\to
XX).
\end{eqnarray}
{}For the $U(3)_{a}\times Sp(1)_{L}\times U(1)_{c}$ D-brane model, $\eta=0,\xi^{2}=1-\xk^{2}$, and all amplitudes involving $X$ or
$A^{3}$ vanish. We obtain
\begin{eqnarray}
{\cal M}(\xg\xg\to gg) & = & \kappa^{2}C_{W}{}^{2}\,{\cal M}(CC\to gg) \, ,
\\
{\cal M}(\xg\xg\to\xg\xg) & = & \xk^{4}C_{W}{}^{4}\,{\cal M}(CC\to CC)+(1-\xk^{2})^{2}C_{W}{}^{4}\,{\cal M}(BB\to BB)\,,\\
{\cal M} (\xg \xg \to Z^0Z^0)  & = & C_W{}^2 S_W{}^2[\xk^4 {\cal M} (C C \to CC)   +(1-\xk^2)^2  {\cal M} (BB \to BB)] \,,\\
{\cal M} (\xg \xg \to W^+W^-)  & = & 0 \, .
\end{eqnarray}
\subsubsection{$\gamma\gamma\to F\bar F$}
Since the vertex operators creating chiral mater fermions contain boundary changing operators connecting two different stacks of intersecting D-branes, say $a$ and $b$, the disk boundary in the amplitudes involving two fermions and two gauge bosons is always attached to two stacks of D-branes. The gauge bosons can couple either to the same stack or to two different stacks. In the latter case, the amplitude with two gauge bosons in the initial state is proportional to $V_s$, which has no poles in the $s$-channel \cite{Lust:2008qc}. The only amplitudes exhibiting $s$-channel poles involve the two initial gauge bosons associated the same stack, but carrying opposite helicities \cite{Lust:2008qc}:
\begin{equation}
{\cal M}(A_{1}^{-},A_{2}^{+},F_{3}^{-},\bar{F}_{4}^{+})=2\, g^{2}\frac{\langle13\rangle^{2}}{\langle32\rangle\langle42\rangle}
\bigg[\frac{t}{s}V_{t}(T^{a_{1}}T^{a_{2}})_{\alpha_{3}\alpha_{4}}+
\frac{u}{s}V_{u}(T^{a_{2}}T^{a_{1}})_{\alpha_{3}\alpha_{4}}\bigg]\ .\label{quarks}\end{equation}
The above equation describes the case of stack $a$, hence the (fermion) spectator indices associated to stack $b$ have been suppressed.
The lowest Regge excitations give rise to the pole term \begin{equation}
{\cal M}(A_{1}^{-},A_{2}^{+},F_{3}^{-},\bar{F}_{4}^{+})\to2\, g^{2}\ {\cal D}^{1234}\frac{\langle13\rangle^{2}}{\langle32\rangle\langle42\rangle}\frac{tu}{M^{2}(s-M^{2})}\ ,\label{qlim}\end{equation}
where the group factor
\be \CD^{1234}\equiv\{T^{a_{1}},T^{a_{2}}\}_{\xa_{3},\xa_{4}}\ .\ee
The group factors and couplings for the processes relevant to $\gamma\gamma\to F\bar F$ are listed in Table~\ref{t3}.
\begin{table}
\caption{Group factors and couplings for the pole terms (\ref{qlim}).}
\begin{tabular}{l|cc}
\hline
\hline
~~ Process~~& ~~Coupling~~ &~~~~$\CD^{1234}$~~~~ \\
\hline
$CC\to q\bar{q}$ & $g_a^2$  & $\frac{1}{3}\delta_{\alpha_{3}\alpha_{4}}$ \\ \hline
$XX\to q_{L}\bar{q}_{R}$ & $g_b^2$ &$\frac{1}{2}$\\
$A^{3}A^{3}\to q_{L}\bar{q}_{R}$ & $g_b^2$ & $\frac{1}{2}$\\
$A^{3}X\to u_{L}\bar{u}_{R}$ &$g_b^2$ & $\frac{1}{2}$\\
$A^{3}X\to d_{L}\bar{d}_{R}$ &$g_b^2$ & $-\frac{1}{2}$\\ \hline
$BB\to q_{R}\bar{q}_{L}$ & $2g_c^2$ & 1\\ \hline
$XX\to e_{R}^{+}e_{L}^{-}$ & $g_b^2$ &$\frac{1}{2}$\\
$A^{3}X\to e_{R}^{+}e_{L}^{-}$ &$g_b^2$ & $-\frac{1}{2}$\\
$A^{3}A^{3}\to e_{R}^{+}e_{L}^{-}$ & $g_b^2$ & $\frac{1}{2}$\\
$XX\to \bar{\nu}_{R}\nu_{L}$ & $g_b^2$ &$\frac{1}{2}$\\
$A^{3}X\to \bar{\nu}_{R}\nu_{L}$ &$g_b^2$ & $\frac{1}{2}$\\
$A^{3}A^{3}\to \bar{\nu}_{R}\nu_{L}$ & $g_b^2$ & $\frac{1}{2}$\\
 \hline
$BB\to e_{R}^{+}e_{L}^{-}$ & $2g_c^2$ & 1\\
$BB\to e_{L}^{+}e_{R}^{-}$ & $2g_c^2$ & 2\\
$BB\to \bar{\nu}_{R}\nu_{L}$ & $2g_c^2$ & 1\\
$BB\to\bar{\nu}_{L}\nu_{R}$ & $2g_c^2$ & 2\\
\hline
\hline
\end{tabular}
\label{t3}
\end{table}

As in the case of $\gamma\gamma$ fusion into gauge boson pairs, the higher level resonances contributing to
$\gamma\gamma\to F\bar F$ come from odd $n$ levels only, so here again, we limit our discussion to $n=1$.
{} In the  $U(3)_{a}\times U(2)_{b}\times U(1)_{c}$ case, the relevant amplitudes
are
\begin{eqnarray}
{\cal M} (\xg \xg \to q_L \bar q_R)  & = & \eta^2 C_W{}^2 \, {\cal M} (X X \to q_L \bar q_R) +S_W{}^2 \, {\cal M} (A^3 A^3 \to q_L \bar q_R)\nonumber \\
 & & + \xk^2 C_W{}^2 \, {\cal M}(CC \to q_L \bar q_R)+~2\eta C_W S_W \, {\cal M}(X A^3 \to q_L \bar q_R) \nn
 & = & (\eta^2 C_W{}^2+S_W{}^2) \, {\cal M} (X X \to q_L \bar q_R) + \xk^2 C_W{}^2 \, {\cal M}(CC \to q_L \bar q_R) \nonumber \\
 & & +~2\eta C_W S_W \, {\cal M}(X A^3 \to q_L \bar q_R) \,, \\[1mm]
{\cal M} (\xg \xg \to q_R \bar q_L)  & = &  \xi^2 C_W{}^2  \, {\cal M} (BB \to q_R \bar q_L) + \xk^2 C_W{}^2 \, {\cal M} (CC \to q_R \bar q_L) \,, \\[1mm]
{\cal M} (\xg \xg \to e_R^+ e_L^-) &  = &  \eta^2 C_W{}^2 \, {\cal M} (X X \to e_R^+ e_L^-)+S_W{}^2 \, {\cal M} (A^3 A^3 \to e_R^+ e_L^-)  \nonumber \\
 &  & + \xi^2 C_W{}^2 \,  {\cal M} (BB \to e_R^+ e_L^-)+~2\eta C_W S_W \, {\cal M}(X A^3 \to e_R^+ e_L^-) \nn
&  = &  (\eta^2 C_W{}^2+S_W{}^2) \, {\cal M} (X X \to e_R^+ e_L^-) + \xi^2 C_W{}^2 \,  {\cal M} (BB \to e_R^+ e_L^-) \nonumber \\
 &  & +~2\eta C_W S_W \, {\cal M}(X A^3 \to e_R^+ e_L^-) \, ,\label{fantif1}\\[1mm]
\CM(\xg \xg \to e_L^+ e_R^-) & = & \xi^2 C_W{}^2\CM(B B \to e_L^+ e_R^-) .\label{fantif2}
\end{eqnarray}
The amplitudes describing neutrino-antineutrino pair production can be obtained from Eqs.(\ref{fantif1}) and (\ref{fantif2}) by the replacement $e^{-}_L\to\nu_L,~ e^{+}_R\to\bar{\nu}_R$.
{}For the $U(3)_{a}\times Sp(1)_{L}\times U(1)_{c}$ D-brane model, we obtain:
\begin{eqnarray}
{\cal M}(\xg\xg\to q_{L}\bar{q}_{R}) & = & \xk^{2}C_{W}{}^{2}\,{\cal M}(CC\to q_{L}\bar{q}_{R})\,,\\[1mm]
{\cal M}(\xg\xg\to q_{R}\bar{q}_{L}) & = & (1-\xk^{2})C_{W}{}^{2}\,{\cal M}(BB\to q_{R}\bar{q}_{L})+\xk^{2}C_{W}{}^{2}\,{\cal M}(CC\to q_{R}\bar{q}_{L})\,,\\[1mm]
\label{41}
{\cal M}(\xg\xg\to e^{\pm}e^{\mp}) & = & (1-\xk^{2})C_{W}{}^{2}\,{\cal M}(BB\to e^{\pm}e^{\mp})\ ,\\[1mm]
{\cal M}(\xg\xg\to \nu\bar{\nu}) & = & (1-\xk^{2})C_{W}{}^{2}\,{\cal M}(BB\to \nu\bar{\nu})\ .
\end{eqnarray}

\subsection{$\bm{e^+e^-}$ annihilation into gauge bosons and resonant contributions to $\bm{e^+e^- \to F \overline F}$}
\subsubsection{$e^+e^-\to\gamma\gamma$, $e^+e^-\to Z^0Z^0$, $e^+e^-\to Z^0\gamma$, $e^+e^-\to W^+W^-$}
Leptons are decoupled from gluons at the disk level because they originate from strings ending on different D-branes. Thus $e^+e^-$ pairs can annihilate into photons and electroweak bosons only.\footnote{$e^+e^- \to \gamma \gamma$ in a toy, one-stack, stringy model has been discussed in~\cite{Cullen:2000ef}.} The corresponding resonance pole terms are obtained by crossing from Eq.(\ref{mhvs}):
 \begin{equation}
{\cal M}([e^{\pm}]_{1}^{-},[e^{\mp}]_{2}^{+},A_{3}^{-},A_{4}^{+},)\to2\, g^{2}\ {\cal D}^{1234}\frac{\langle13\rangle^{2}}
{\langle 14\rangle\langle 24\rangle}\frac{tu}{M^{2}(s-M^{2})}\ ,
\end{equation}
with the same group factors as in Table~\ref{t3}, but running in the time-reversed channels.
In the $U(3)_{a}\times U(2)_{b}\times U(1)_{c}$ case, the physical amplitudes for the processes under consideration are
\begin{align}
{\cal \mathcal{M}}(e_{R}^{+}e_{L}^{-}\rightarrow\gamma\gamma) & =\eta^{2}C_{W}{}^{2}\,{\cal M}(e_{R}^{+}e_{L}^{-}\rightarrow XX)+S_{W}{}^{2}\,{\cal M}(e_{R}^{+}e_{L}^{-}\rightarrow A^{3}A^{3})\nonumber \\
 & \quad+\xi^{2}C_{W}{}^{2}\,{\cal M}(e_{R}^{+}e_{L}^{-}\rightarrow BB)+~2\eta C_{W}S_{W}\,{\cal M}(e_{R}^{+}e_{L}^{-}\rightarrow XA^{3})\nonumber \\
 & =(\eta^{2}C_{W}{}^{2}+S_{W}{}^{2})\,{\cal M}(e_{R}^{+}e_{L}^{-}\rightarrow XX)+\xi^{2}C_{W}{}^{2}\,{\cal M}(e_{R}^{+}e_{L}^{-}\rightarrow BB)\nonumber\\
 & \quad+2\eta C_{W}S_{W}\,{\cal M}(e_{R}^{+}e_{L}^{-}\rightarrow XA^{3})\,,\\[1mm]
\mathcal{M}(e_{L}^{+}e_{R}^{-}\rightarrow\gamma\gamma) & =\xi^{2}C_{W}{}^{2}\mathcal{M}(e_{L}^{+}e_{R}^{-}\rightarrow BB) \,,\\[1mm]
\mathcal{M}(e_{R}^{+}e_{L}^{-}\rightarrow Z^0Z^0)  & =(\eta^{2}S_{W}{}^{2}+C_{W}{}^{2})\,{\cal M}(e_{R}^{+}e_{L}^{-}\rightarrow XX)+\xi^{2}S_{W}{}^{2}\,{\cal M}(e_{R}^{+}e_{L}^{-}\rightarrow BB)\nonumber\\
 & \quad+2\eta C_{W}S_{W}\,{\cal M}(e_{R}^{+}e_{L}^{-}\rightarrow XA^{3})\,,\\[1mm]
\mathcal{M}(e_{L}^{+}e_{R}^{-}\rightarrow Z^0Z^0) & =\xi^{2}S_{W}{}^{2}\mathcal{M}(e_{L}^{+}e_{R}^{-}\rightarrow BB) \,,\\[1mm]
\mathcal{M}(e_{R}^{+}e_{L}^{-}\rightarrow Z^0\gamma) & =S_{W}C_W(\eta^{2}+1)\,{\cal M}(e_{R}^{+}e_{L}^{-}\rightarrow XX)+\xi^{2}S_{W}{}C_W\,{\cal M}(e_{R}^{+}e_{L}^{-}\rightarrow BB)\nonumber\\
 & \quad+\eta (C_{W}{}^2+S_{W}{}^2)\,{\cal M}(e_{R}^{+}e_{L}^{-}\rightarrow XA^{3})\,,\\[1mm]
\mathcal{M}(e_{L}^{+}e_{R}^{-}\rightarrow Z^0\gamma) & =\xi^{2}S_{W}C_W\mathcal{M}(e_{L}^{+}e_{R}^{-}\rightarrow BB) \,,\\[1mm]
\mathcal{M}(e_{R}^{+}e_{L}^{-}\rightarrow W^+W^-) &= {\cal M}(e_{R}^{+}e_{L}^{-}\rightarrow A^{3}A^{3}) \, ,\\[1mm]
\mathcal{M}(e_{L}^{+}e_{R}^{-}\rightarrow W^+W^-) & = 0 \, .
\end{align}
{}For the $U(3)_{a}\times Sp(1)_{L}\times U(1)_{c}$ D-brane model,
we have
 \begin{align}
{\cal \mathcal{M}}(e_{R}^{+}e_{L}^{-}\rightarrow\gamma\gamma) & =\xi^{2}C_{W}{}^{2}\,{\cal M}(e_{R}^{+}e_{L}^{-}\rightarrow BB) \,,\\[1mm]
\mathcal{M}(e_{L}^{+}e_{R}^{-}\rightarrow\gamma\gamma) & =\xi^{2}C_{W}{}^{2}\mathcal{M}(e_{L}^{+}e_{R}^{-}\rightarrow BB) \,,\\[1mm]
\mathcal{M}(e_{R}^{+}e_{L}^{-}\rightarrow Z^0Z^0) & = \xi^{2}S_{W}{}^{2}{\cal M}(e_{R}^{+}e_{L}^{-}\rightarrow BB)\,,\\[1mm]
\mathcal{M}(e_{L}^{+}e_{R}^{-}\rightarrow Z^0Z^0)& =\xi^{2}S_{W}{}^{2}\mathcal{M}(e_{L}^{+}e_{R}^{-}\rightarrow BB) \,,  \\[1mm]
\mathcal{M}(e_{R}^{+}e_{L}^{-}\rightarrow Z^0\gamma) & = \xi^{2}S_{W}{}C_W\,{\cal M}(e_{R}^{+}e_{L}^{-}\rightarrow BB) \,,\\[1mm]
\mathcal{M}(e_{L}^{+}e_{R}^{-}\rightarrow Z^0\gamma) &=\xi^{2}S_{W}C_W\mathcal{M}(e_{L}^{+}e_{R}^{-}\rightarrow BB) \,, \\[1mm]
\mathcal{M}(e_{R}^{+}e_{L}^{-}\rightarrow W^+W^-) & = {\cal M}(e_{R}^{+}e_{L}^{-}\rightarrow A^{3}A^{3}) \,,\\[1mm]
\mathcal{M}(e_{L}^{+}e_{R}^{-}\rightarrow W^+W^-) & = 0 \, .
\end{align}
\subsubsection{Resonant contributions to $e^+e^-\to e^+e^-$, $e^+e^-\to \nu\bar{\nu}$,
$e^+e^-\to q\bar{q}$}
\label{eeee}

Four-fermion amplitudes \cite{Lust:2008qc} are not universal -- they depend on the internal radii and other details of extra dimensions already at the disk level. In particular, they contain resonance poles due to Kaluza-Klein excitations. More serious problems though are due to the presence of resonance poles associated to both massless and massive particles that are either unacceptable from the phenomenological point of view, or are expected to receive large mass corrections
due to quantum (anomaly) effects, see Ref.\cite{Anchordoqui:2009mm} for more details. {}For example, the same Green-Schwarz mechanism that generates non-zero masses for anomalous gauge bosons does also affect the masses of their Regge excitations.
{}For the above reasons, phenomenological  analysis of $e^+e^-$ annihilation into lepton-antilepton pairs will be quite complicated, as described in more detail in the following Sec.~IV B.

Here, we focus on the lowest Regge excitations of the photon and $Z^0$, remaining in the spectrum of any realistic model. Since we are considering energies far above the electroweak scale, we can replace $\gamma$ and $Z^0$ by the neutral gauge bosons of unbroken $SU(2)\times U(1)_Y$.

At the lowest, $n=1$ level, each gauge boson comes with several Regge excitations with spins ranging from 0 to 2, but only two particles couple to quark-antiquark and lepton-antilepton pairs: one spin 2 boson and one spin 1 vector particle \cite{Anchordoqui:2008hi}. All three-particle couplings involving one Regge excitation, one fermion and one antifermion have been determined in Ref.\cite{Anchordoqui:2008hi} by using the factorization methods. These S-matrix elements are completely sufficient for
reconstructing the resonance part of four-fermion amplitudes \cite{Anchordoqui:2008hi} by using the Wigner matrix techniques. In the center of mass frame, the relevant amplitudes can be written as
\begin{eqnarray}
{\cal M}(e^-_Le^+_R\to F_L\bar F_R) & \to &
\frac{M^2}{s-M^2}\frac{e^2}{4}\Big(\frac{Y_F}{C_W^2}+\frac{I_{3F}}{S_W^2}\Big)\Big[d^2_{1,1}(\theta)+
\frac{1}{3}d^1_{1,1}(\theta)\Big],\\[1mm]
{\cal M}(e^-_Le^+_R\to F_R\bar F_L) & \to & \frac{M^2}{s-M^2}\frac{e^2}{4}\frac{Y_F}{C_W^2}
\Big[d^2_{1,-1}(\theta)+
\frac{1}{3}d^1_{1,-1}(\theta)\Big],\\[1mm]
{\cal M}(e^-_Re^+_L\to F_L\bar F_R) & \to & \frac{M^2}{s-M^2}\frac{e^2}{2}\frac{Y_F}{C_W^2}
\Big[d^2_{1,-1}(\theta)+
\frac{1}{3}d^1_{1,-1}(\theta)\Big],\\[1mm]
{\cal M}(e^-_Re^+_L\to F_R\bar F_L) & \to & \frac{M^2}{s-M^2}\frac{e^2}{2}\frac{Y_F}{C_W^2}
\Big[d^2_{1,1}(\theta)+
\frac{1}{3}d^1_{1, 1}(\theta)\Big],
\end{eqnarray}
where $Y_F$ is the fermion hypercharge, $I_{3F}$ is the fermion weak isospin, and
\begin{equation}
d^2_{1,\pm 1}(\theta)=\frac{1\pm\cos\theta}{2}(2\cos\theta\mp 1)\ ,
\qquad\quad d^1_{1,\pm 1}(\theta)=\frac{1\pm\cos\theta}{2}\ ,
\end{equation}
are the spin 2 and spin 1 Wigner matrix elements~\cite{WignerGT,EdmondsAM}, respectively.
A very  interesting aspect of the above result is that string theory predicts
the {\em  precise} value, equal 1/3, of the relative weight of spin 2 and spin 1 contributions.

Here again, we would like to stress that although the full four-fermion scattering amplitudes are model-dependent, their resonance parts are universal because
the three-particle couplings involving one Regge excitation and two massless particles do not depend on the compactification space \cite{Feng:2010yx}.

\section{Phenomenology}
\label{IV}

In this section we study the distinct phenomenology of Regge recurrences
arising in the $\gamma \gamma$ and $e^+ e^-$ beam settings.

\subsection{$\bm{\gamma \gamma}$ collisions}

As an illustration of the CLIC potential to uncover string signals,
we focus attention on dominant $\gamma \gamma \to e^+ e^-$
scattering, within the context of the $U(3)_a \times Sp(1)_L \times
U(1)_c$ D-brane model. Let us first isolate the contribution to the
partonic cross section from the first resonant state, $B^*$. The $s$-channel
pole term of the average square amplitude can be obtained from the
formula (\ref{41}) by taking into account all possible initial
polarization configurations. However, for phenomenological purposes,
the pole needs to be softened to a Breit-Wigner form by obtaining and
utilizing the correct {\em total} widths of the resonance. After this
is done we obtain
\begin{eqnarray}
  |{\cal M} (\xg\xg \to  e^+ e^-)|^2 & = & (1 + 4 ) \, (1-\xk^2)^2 \, C_W^4  \, \frac{4 g_c^4}{M^4}\
 \left [\frac{  u   t (   u^2+   t^2)}{( s-M^2)^2 +
(\Gamma_{B^*}^{J=2}\ M)^2} \right] \, ,
\label{gagaee}
\end{eqnarray}
where the factor of $(1+4)$  in the numerator accounts for the fact that the $U(1)_c$ charge of $e_R$ is twice that of $e_L$. The decay width of $B^*$ is given by \ba \xG_{B^*}^{J=2} & = &
\xG_{B^* \to l \bar l}^{J=2} + \xG_{B^* \to q_R \bar q_L}^{J=2} +
\xG_{B^* \to B B}^{J=2} \nn & = & \frac {g_c^2} {\pi} M \lsb \frac
{1} {40}\lb \frac 5 2 N_e+ \frac 1 2 N_\nu + \frac 1 2 N_q\rb + \frac
1 {5 N} \rsb\nn & = & \frac {13}{20} \frac {g_c^2} {4\pi} M, \ea
where $N_e = 3, N_\nu = 3, N_q = 18$. The first term comprises the
contribution from the left-handed ($N_e/2$) and right-handed ($2 N_e$)
electrons, the second term ($N_\nu/2$) comes from the left-handed
neutrinos, and the third term $(N_q/2)$ subsume the right-handed
quarks.

The total cross section at an $e^{+}e^{-}$ linear collider can be obtained by folding $\hat{\sigma} (\hat{s})$ with the photon distribution function~\cite{Jikia:1991hc} \begin{equation}
  \sigma_{\rm tot}(e^+e^- \Rightarrow \gamma\gamma \to  e^+ e^-) =\int^{x_{\rm max}}_{M/\sqrt{s}} dz \ \frac{d{\cal
      L}_{\gamma\gamma}}{dz} \ \hat{\sigma}(\hat{s}=z^2 s ) \,,
 \end{equation}
where  $\hat{s}$ and $s$ indicate respectively the
center-of-mass energies of the $\gamma\gamma$ and the parent $e^{+}e^{-}$
systems and
\begin{eqnarray}
\frac{d{\cal L}_{\gamma\gamma}}{dz}=2z\int_{z^2/x_{\rm max}}^{x_{\rm max}}
 \frac{dx}{x} f_{\gamma/e}(x)f_{\gamma/e}(z^2/x) \,
\end{eqnarray}
is the distribution function of photon luminosity. The energy spectrum of the back scattered photon in unpolarized incoming $e \gamma$ scattering is given by
\begin{eqnarray}
\label{structure}
f_{\gamma/e}(x)=\frac{1}{D(\xi)}\left[1-x+\frac{1}{1-x}-
\frac{4x}{\xi(1-x)}+\frac{4x^{2}}{\xi^{2}(1-x)^2}\right],~~~(x<x_{\rm max}) \,,
\end{eqnarray}
where $x=2
\omega/\sqrt{s}$ is the fraction of the energy of the incident electron carried
by the back-scattered photon  and $x_{\rm max}=2
\omega_{\rm max}/\sqrt{s}=\xi/(1+\xi)$. For $x>x_{\rm max}$,
$f_{\gamma/e}=0$. The function $D(\xi)$ is defined as
\begin{equation}
D(\xi)= \left(1-\frac{4}{\xi}-\frac{8}{\xi^{2}} \right)\ln(1+\xi)+\frac{1}{2}+\frac{8}{\xi}-\frac{1}{2(1+\xi)^{2}}.
\end{equation}
where $\xi=2 \omega_0\sqrt{s}/{m_e}^2$, $m_{e}$ and $\omega_0$ are respectively the electron mass and laser-photon energy, and (of course) the incoming electron energy is $\sqrt{s}/2$. In our evaluation, we choose $\omega_0$ such that it maximizes the backscattered photon energy without spoiling the luminosity through $e^{+}e^{-}$ pair creation, yielding ${\xi}=2(1+\sqrt{2})$, $x_{\rm
  max}\simeq 0.83$ and $D(\xi) \approx 1.84$~\cite{Cheung:1992jn}.

\begin{figure}[tbp]
\postscript{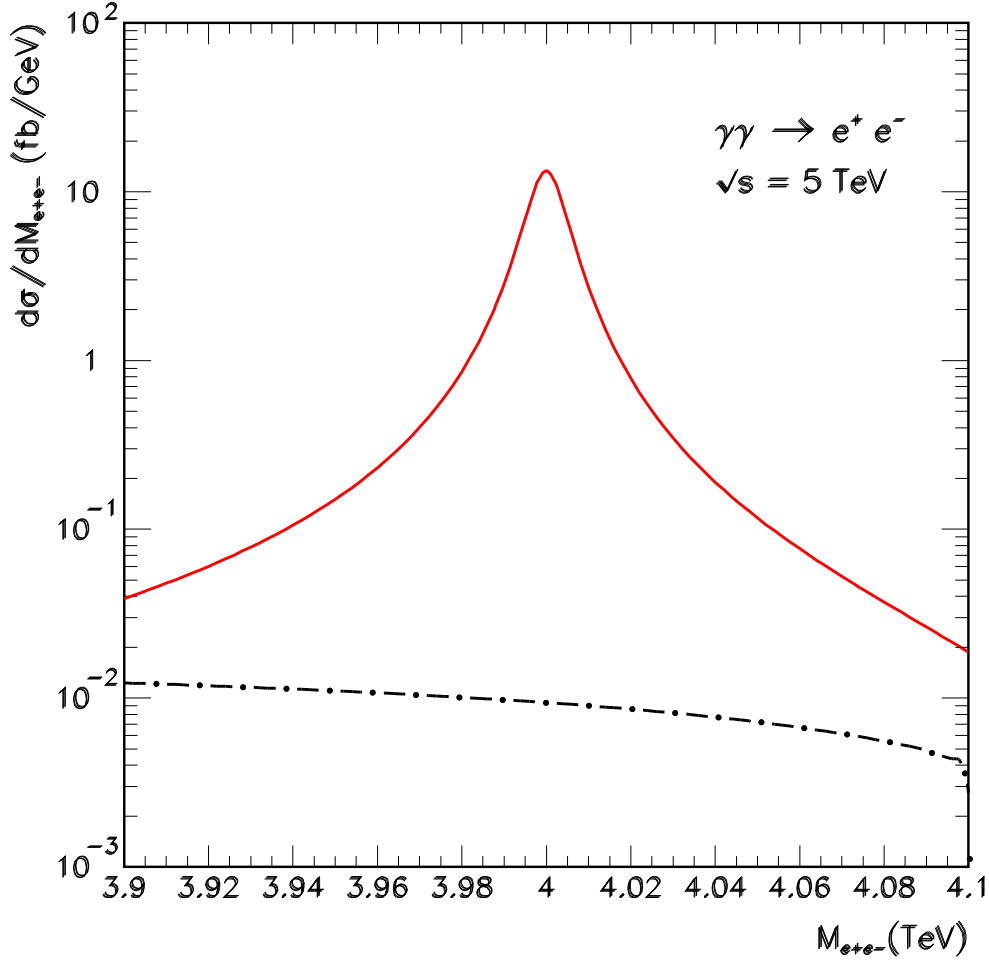}{0.8}
\caption{$d\sigma/dM_{e^+e^-}$ (units of fb/GeV) {\em vs.}
  $M_{e^+e^-}$ (TeV) is plotted for the case of SM background (dot-dashed line)
  and (first resonance) string signal + background (solid line), for  $M=4$~TeV and $\sqrt{s} = 5$~TeV. (We have taken $\kappa = 0.14.$)}
\label{fig:bump}
\end{figure}

We study the signal-to-noise of Regge excitations in data binned
according to the invariant mass $M_{e^+e^-}$ of the $e^+ e^-$ pair, after
setting cuts on the different electron-positron rapidities, $|y_1|, \, |y_2|
\le 2.4$ and transverse momenta $p_{\rm
  T}^{1,2}>50$~GeV.  With the definitions $Y\equiv \frac 1 2 (y_1 +
y_2)$ and $y \equiv \frac 1 2 (y_1-y_2)$, the cross section per
interval of $M_{e^+e^-}$ for $\xg\xg \rightarrow e^+ e^-$ is given
by \begin{eqnarray} \frac{d\sigma}{dM_{e^+e^-}} & = & \sqrt s z^{3}\ \left[
    \int_{-Y_{\rm max}}^{0} dY \ f_{\xg/e} (x_a) \right. \ f_{\xg/e}
  (x_b) \ \int_{-(y_{\rm max} + Y)}^{y_{\rm max} + Y} dy
  \left. \frac{d\hat \sigma}{d\hat t}\right|_{\xg\xg\rightarrow
    e^+e^-}\ \frac{1}{\cosh^2
    y} \nonumber \\
  & + &\int_{0}^{Y_{\rm max}} dY \ f_{\xg/e} (x_a) \ f_{\xg/e} (x_b) \
  \int_{-(y_{\rm max} - Y)}^{y_{\rm max} - Y} dy
  \left. \left. \frac{d\hat \sigma}{d\hat t}\right|_{\xg\xg\rightarrow
      e^+ e^-}\ \frac{1}{\cosh^2 y}
  \right] \label{longBH} \end{eqnarray} where $z^2 = M_{e^=e^-}^2/s$, $x_a =z
e^{Y}$, $x_b =z e^{-Y},$ and \begin{equation} |{\cal M}(\xg\xg \to
  e^+e^-) |^2 = 16 \pi \hat s^2 \, \left. \frac{d\sigma}{d\hat t}
  \right|_{\xg \xg \to e^+ e^-} \, .
\end{equation}
The string signal is
calculated using (\ref{longBH}) with the corresponding $\gamma \gamma \to e^+ e^-$ scattering amplitude given in Eq.~(\ref{gagaee}). The SM background is calculated using
\be
\frac {d \hat \xs}{d \hat t} =\frac {2 \pi
\xa^2}{\hat s^2} \lb\frac {\hat u}{\hat
t} + \frac {\hat t}{\hat u}\rb \, .
\label{gagaeeSM}
\ee The kinematics of the scattering also provides the relation $M_{e^+e^-} = 2p_T \cosh y$, which when combined with the standard cut $p_T \agt p_{T,\rm min}$, imposes a {\em lower} bound on $y$ to be implemented in the limits of integration.  The $Y$ integration range in Eq.~(\ref{longBH}), $Y_{\rm max} = {\rm min} \{ \ln(x_{\rm{max}}/z),\ \ y_{\rm max}\}$, comes from requiring $x_a, \, x_b < x_{\rm{max}}$ together with the rapidity cuts $0 <|y_1|, \, |y_2| < 2.4$. Finally, the Mandelstam invariants occurring in the cross section are given by $\hat s = M_{e^+e^-}^2,$ $\hat t = -\frac 1 2 M_{e^+e^-}^2\ e^{-y}/ \cosh y,$ and $\hat u = -\frac 1 2 M_{e^+e^-}^2\ e^{+y}/ \cosh y.$ In Fig.~\ref{fig:bump} we show a representative plot of the invariant mass spectrum, for $M = 4$~TeV and $\sqrt{s} = 5$~TeV.

We now estimate (at the parton level)  the signal-to-noise ratio at CLIC. Standard bump-hunting methods, such as obtaining cumulative cross sections, $\sigma (M_0) = \int_{M_0}^\infty \frac{d\sigma}{dM_{e^+e^-}} \, \, dM_{e^+e^-}$, from the data and searching for regions with significant deviations from the SM background, may reveal an interval of $M_{e^+e^-}$ suspected of containing a bump.  With the establishment of such a region, one may calculate the detection significance
\begin{equation}
  S_{\rm det} = \frac{N_{\rm S}}{\sqrt{N_{\rm B} + N_{\rm S}}} \, ,
\end{equation}
with the signal rate $N_{\rm S}$ estimated in the invariant mass window $[M - 2 \Gamma, \, M + 2 \Gamma]$, and the number of background events $N_{\rm B}$ defined in the same $e^+e^-$ mass interval for the same integrated luminosity~\cite{Anchordoqui:2006pb}.  For $\sqrt{s} = 5$~TeV and $M_s = 4$~TeV we expect $S_{\rm det} \simeq 139/12 = 11\sigma$, after the first fb$^{-1}$ of data collection.  The spin-2  nature of  $\gamma \gamma \to e^+ e^-$ Regge recurrences would make them smoking guns for low mass scale D-brane string compactifications.

\subsection{$\bm{e^+ e^-}$ collisions}

\begin{figure}[tbp]
\postscript{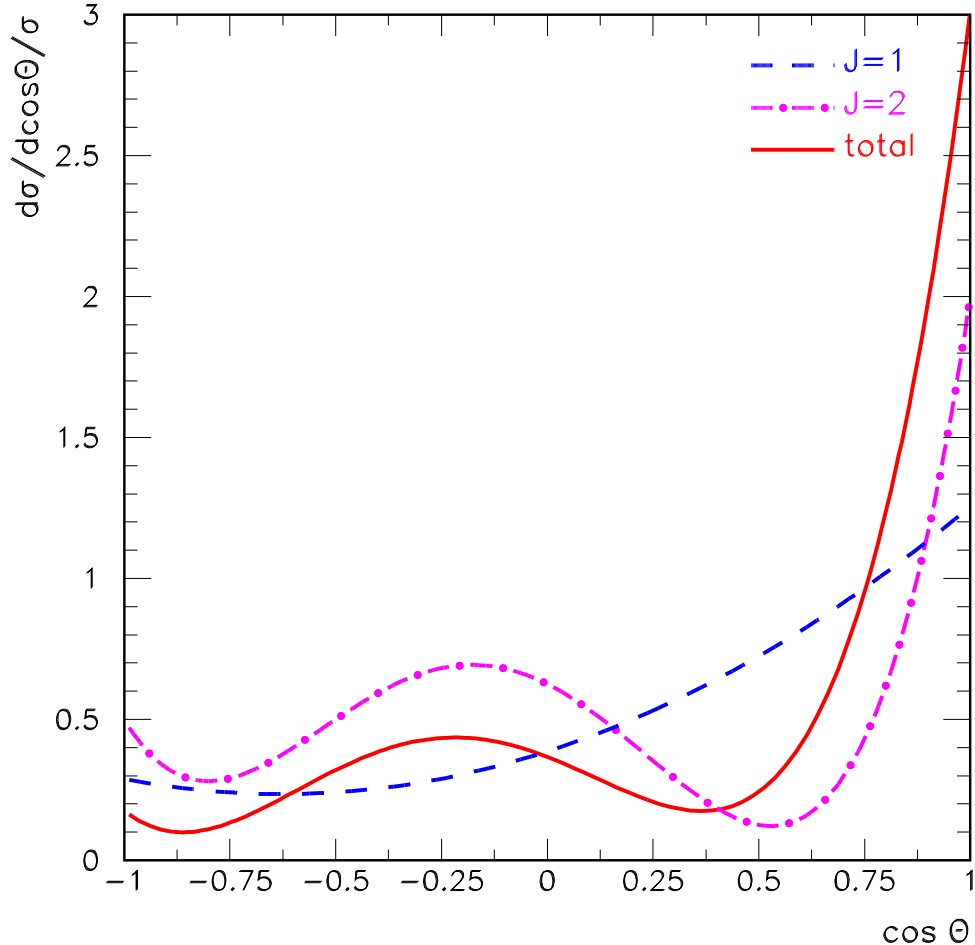}{0.8}
\caption{Normalized angular distributions of Regge recurrences with spin 1, 2, and total in the $e^+ e^- \to \mu^+ \mu^-$ channel.}
\label{fig:ad}
\end{figure}

We assume that the $e^+e^-$ center-of-mass energy will be tuned to
contain the interesting range highlighted by LHC data and that the
resolution of the machine will be sufficient to probe narrow
resonances. We are interested in the $e^+e^-$ annihilation into lepton-antilepton pairs, in particular in $e^-e^+ \to \mu^-\mu^+$. Phenomenological analysis of such processes will be quite complicated, due the presence of model-dependent backgrounds of Kaluza-Klein (KK) excitations, anomalous gauge gauge bosons and their Regge excitations. Weakly-interacting KK excitations are expected to have masses lower than the string scale \cite{Anchordoqui:2009mm}, and can appear as resonances in the $e^+e^-$ annihilation channel. Their signals will be similar to a generic $Z'$, with a unique angular momenta, commonly $J=1$ and will not provide direct evidence for the superstring substructure. The signals of gauge bosons associated to anomalous $U(1)$ gauge bosons, with masses always lower than the string scale, varying from a loop factor to a large suppression by the volume of the bulk~\cite{Antoniadis:2002cs}, will have a similar character. We assume that no accidental degeneracy occurs between these particles and Regge excitations, so that the string signal discussed Sec.~\ref{eeee} can be safely isolated from the background. Even in this case, however, there is a certain amount of ambiguity due to the presence of Regge excitations of anomalous $U(1)$'s with masses shifted by radiative corrections \cite{Kitazawa:2010gh}. If this shift is large, there will be a separate resonance peak, but if it is small, it will affect the normalization of the signal.

\begin{figure}[tbp]
\postscript{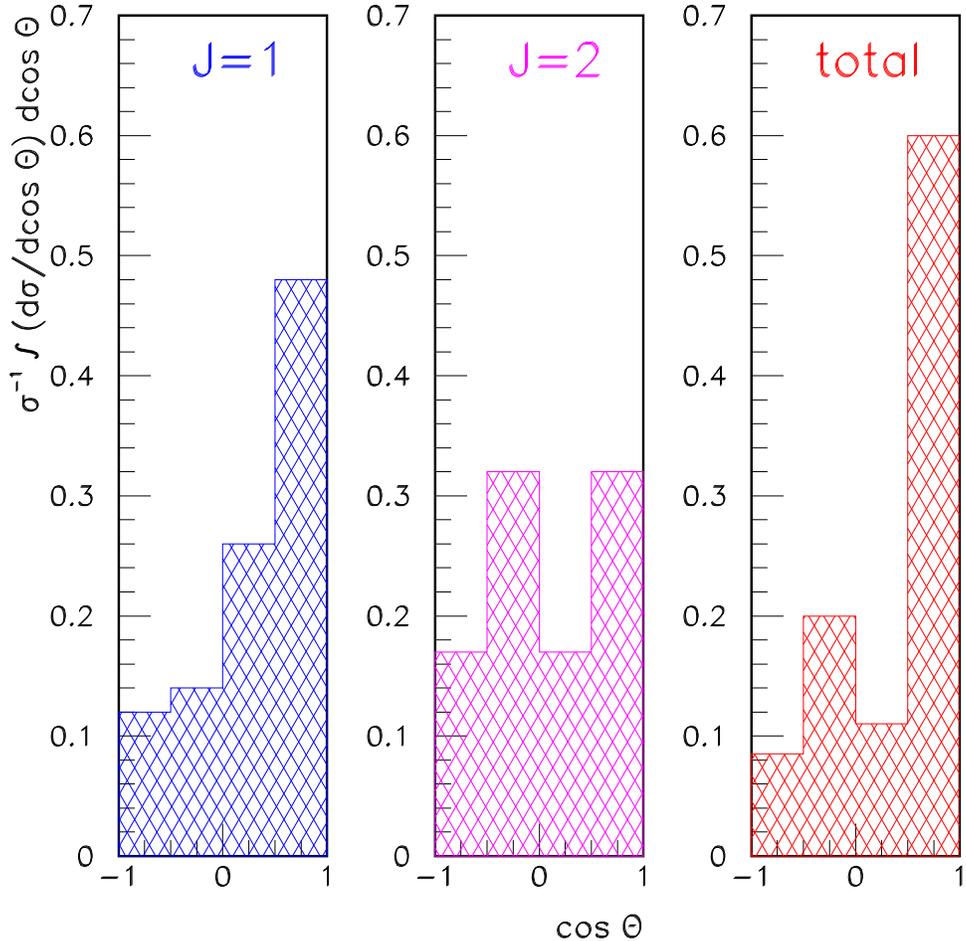}{0.8}
\caption{Binned angular distributions of Regge recurrences with spin 1, 2, and total in the $e^+ e^- \to \mu^+ \mu^-$ channel.}
\label{fig:3}
\end{figure}

Should a resonance be found, a strong discriminator
between models will be the observed angular distribution. Typical
candidates for new physics such as $Z'$ will have a unique angular
momenta, commonly $J=1$. It is an interesting and exciting peculiarity
of Regge recurrences that the angular momenta content of the energy
state is more complicated. As we have shown in Sec.~\ref{eeee}, for
the lightest Regge excitation there is a specific combination of $J=1$
and $J=2$, which are access by the $e^+ e^-$ beam setting.
Specializing at this point to $e^-e^+ \to \mu^-\mu^+$, so that $I_{3F_L}
= Y_{F_L} = \frac{1}{2} Y_{F_R}= -1/2$, we obtain the normalized angular distribution
\begin{equation}
\frac{d\sigma/d\cos{\theta}}{\sigma}  =
{\cal N}\ \left\{\left[4 +\left(\frac{1}{2\ S_W^2}\right)^2 \right]
D_+(\theta)^2 + 2 \ D_-(\theta)^2\right\}\ \ ,
\end{equation}
where
\begin{equation}
D_\pm (\theta) \equiv d^2_{1,\pm 1}(\theta) + \frac{1}{3}\ d^1_{1, \pm 1}(\theta)\
\end{equation}
and
\begin{equation}
{\cal N}^{-1} =  (64/135) \left[6 + \left(\frac{1}{2\ S_W^2}\right)^2\right] 
\, . 
\end{equation}
For the $J=2$ piece alone, the normalization constant is
\begin{equation}
{\cal N}_2^{-1} =  (2/5) \left[6 + \left(\frac{1}{2\ S_W^2}\right)^2\right]
\end{equation}
whereas for the $J=1$ piece alone, the normalization constant is
\begin{equation}
{\cal N}_1^{-1} = (2/27) \left[6 + \left(\frac{1}{2\ S_W^2}\right)^2\right] \, .
\end{equation}
In Fig.~\ref{fig:ad} we show the resulting angular distributions. The predicted dimuon angular distribution has a pronounced forward-backward asymmetry. This is a realistic target for CLIC searches of low-mass scale string theory signals. (Note that the $e^+e^- \to e^+ e^-$ Coulomb scattering background, which peaks in the forward direction, tends to wash out the predicted string signal.) In Fig.~\ref{fig:3} we show the binned angular distributions. It is clearly seen that  it would be easy to distinguish the string excitation from single $J=2$ resonance in the dimuon angular distribution. To  completely isolate the Regge excitation from a $J=1$ resonance, one can use string predictions in alternative channels, e.g. $\gamma \gamma \to e^+ e^-$.

\section{Conclusions}
\label{V}

In this paper, we have explored the discovery potential of the proposed $e^+e^-$ and $\gamma\gamma$ colliders to unmask string resonances. We  have studied the direct production of Regge excitations, focusing on the first excited level of open strings localized on the worldvolume of D-branes. In such a D-brane construction the resonant parts of the relevant string theory amplitudes are {\it universal} to leading order in the gauge coupling. Therefore, it is feasible to extract genuine string effects which are independent of the compactification scheme. Among the various processes, we found that the $\gamma \gamma \to e^+ e^-$ scattering proceeds only through a spin-2 Regge state.  Our detailed phenomenological studies suggest that for this specific channel, string scales as high as 4~TeV can  be unmasked at the 11$\sigma$ level with the first fb$^{-1}$ of data collected at $\sqrt{s} \approx 5$~TeV. We have also investigated intermediate Regge states of $e^+e^- \to F \bar F$ and we have shown that string theory predicts the {\em precise} value, equal 1/3, of the relative weight of spin 2 and spin 1 contributions. The potential benefit of this striking result becomes evident when analyzing the dimuon angular distribution, which has a pronounced forward-backward asymmetry, providing a very distinct signal of the underlying string physics.

\section*{Acknowledgements}
We are very grateful to Maria Krawczyk for her encouragement to pursue this project.
L.A.A.\ is supported by the U.S. National Science Foundation (NSF)
Grant No PHY-0757598, and the UWM Research Growth Initiative.  W-Z.F.\ , H.G.\
and  T.R.T.\
are supported by the U.S. NSF Grant PHY-0757959.  Any
opinions, findings, and conclusions or recommendations expressed in
this material are those of the authors and do not necessarily reflect
the views of the National Science Foundation.

\end{document}